\documentclass[fleqn,usenatbib]{mnras}
\DeclareMathAlphabet{\mathpzc}{OT1}{pzc}{m}{it}
\twocolumn
\usepackage{mathptmx}

\usepackage[T1]{fontenc}
\usepackage{ae,aecompl, footnote}
 
\usepackage{graphicx}	
\usepackage{amsmath}	
\usepackage{amssymb}	
\usepackage{bm}
\usepackage{xcolor}
\usepackage{wrapfig}
\usepackage{hyperref}

\title[Wasserstein distance as a new tool]{Wasserstein distance as a new tool for discriminating cosmologies through the topology of large scale structure}

\author[M. Tsizh, V. Tymchyshyn and F. Vazza]{
Maksym Tsizh$^{1,2}$,\thanks{corresponding author: maksym.tsizh@lnu.edu.ua}
Vitalii Tymchyshyn$^{3,4}$,
Franco Vazza$^{2,5,6}$\\
$^1$Astronomical Observatory of Ivan Franko National University of Lviv, Kyryla i Methodia str. 8, Lviv, 
79005, Ukraine \\
$^{2}$Dipartimento di Fisica e Astronomia, Universitá di Bologna, Via Gobetti 92/3, 40121, Bologna, Italy \\
$^3$Synergetics department, Bogolyubov Institute for Theoretical Physics, Metrolohichna 14-b, UA-02000 Kyiv, Ukraine,\\
$^4$Data Science Center, Kyiv Academic University, UA-03142 Kyiv, Ukraine\\
$^{5}$Institute of Radioastronomy - INAF, Via Gobetti 101, 40129 Bologna, Italy\\
$^{6}$Hamburger Sternwarte, Gojenbergsweg 112, 21029 Hamburg, Germany}
\date{10 May 2023}

\pubyear{2023}
\begin{document}
\label{firstpage}
\pagerange{\pageref{firstpage}--\pageref{lastpage}}
\maketitle

\begin{abstract}
In this work we test Wasserstein distance in conjunction with persistent homology, as a tool for discriminating large scale structures of simulated universes with different values of $\sigma_8$ cosmological parameter (present root-mean-square matter fluctuation averaged over a sphere of radius 8 Mpc comoving). The Wasserstein distance (a.k.a. the pair-matching distance) was proposed to measure the difference between two networks in terms of persistent homology. The advantage of this approach consists in its non-parametric way of probing the topology of the Cosmic web, in contrast to graph-theoretical approach depending on linking length. By treating the halos of the Cosmic Web as points in a point cloud we calculate persistent homologies, build persistence (birth-death) diagrams and evaluate Wasserstein distance between them. The latter showed itself as a convenient tool to compare simulated Cosmic webs. We show that one can discern two Cosmic webs (simulated or real) with different $\sigma_8$ parameter. It turns out that Wasserstein distance's discrimination ability depends on redshift $z$, as well as on the dimensionality of considered homology features. We find that the highest discriminating power this tool obtains at $z=2$ snapshots, among the considered $z=2$, $1$, and $0.1$ ones.
\end{abstract}


\section{Introduction}
The quantitative exploration of the large scale distribution of matter as a complex web has been consolidated in this century. The notion of the Cosmic Web, traced by matter halos and the galaxies they contain, has emerged and confidently established in the humankind's view on the Universe, dubbing the complex large scale distribution of matter.

The Cosmic Web is often observed  using galaxies as pinpoint for the underlying matter distribution, thanks to large optical surveys such as Sloan Digital Sky Survey \citep[SDSS,][]{Tegmark04}, 2 Micron All-Sky Survey \citep[2MASS,][]{Huchra12}, and VIMOS Public Extragalactic Redshift Survey \citep[VIPERS,][]{Guzzo14}.
On the other hand, only in recent years, signatures from the diffuse gas in  filaments have been reported with stacking techniques in the X-ray \citep[][]{2020A&A...643L...2T}, in the microwaves through the Sunyaev Zeldovich effect \citep[][]{2019A&A...624A..48D} and at radio frequencies \citep[][]{vern21}.

The many potential large-scale correlations between observable quantities and the intrinsic topological and morphological properties of the Cosmic Web call for a robust geometrical and topological characterization of its global network.
Many numerical algorithms have been developed over the years in order to capture and describe the complex hierarchy of structures in the matter web of cosmological simulations \citep[e.g.,][for  reviews]{2014MNRAS.441.2923C, Libeskind18}.
 
The complex networks approach has also found its application in large-scale analysis. It treats halos or galaxies as the vertices of a complex network (graph) and exploits the network characteristics (metrics) to shed light on the nature of the Cosmic Web. For example, it can be used to determine the type of structure to which a halo belongs \citep{Tsizh2020}, or it can relate network metrics with observable quantities of galaxies of the Cosmic Web \citep{deRegt2018}. Network analysis also makes it possible to quantitatively compare the degree of self-organization and complexity of the architectures of entirely dissimilar systems, like the Cosmic Web and the human brain \citep[e.g.][]{brain}.
The neighboring method of analysis, typical for graph-(network-)like data, is topological data analysis that started to conquer its place in the discussed field. In particular, its subsection called \textit{persistent homology} will be of interest to this work.

Persistent homology is characterized by a set of suitable tools to process the structures of the Cosmic Web, which has been recently explored by a few works.
The first introduction of persistent homology into Cosmic Web science was probably made in 2011 by \citet{Weygaert2011}.
During the next decade, this group  did systematic work studying persistence (birth-death) diagrams, Betti numbers and curves, persistence curves, and other instruments for the Cosmic Web. The same group has recently studied the evolution of persistence of homologies in $\Lambda$CDM  simulations \citep[e.g.][]{Wilding2021}. The follow-up work discovered the multiscale nature of the simulated Cosmic Web by estimating the persistence of topological features (holes) of different dimensions at different scales \citep[][]{Bermejo2022}. While the aforementioned works explored the results of the cosmological simulations, the real data of SDSS catalog was probed with this approach in \citet{Kimura2017}, where the authors introduced the idea of measuring distances between persistent diagrams of large-scale structures. The persistent voids and filaments search in SDSS was conducted in \citet{Xu2019}, while the analysis of  aperture masses obtained from cosmic shear as a result of gravitational lensing has been performed with persistent homology to constrain cosmological parameters, in recent works by \citet{Heydenreich2021} and  \citet{Heydenreich2022}. 

The persistent homology approach can be also applied to continuous reconstructions of the large scale structure data, as was shown in \citet{Cisewski2014}, where the authors analyzed persistent homology of the H-I density field in the intergalactic medium, using the Lyman-alpha forest data. Similarly, the continuous field of reionization bubbles has also become the object of persistent homologies studies in \citet{Elbers2019} and \citet{Elbers2022}.
Finally, the simulation of the interstellar medium together with its magnetic field was subjected to topological data analysis in \citet{Makarenko2018}.

Another well-known way of exploring the large scale structure of the Universe is by comparing the distribution of perturbations to the Gaussian random field. Topology data analysis and persistent homology in particular, manifest themselves as a suitable instrument for this problem, as can be seen from \cite{Feldbrugge2019}, \cite{Biagetti2021}, \cite{Pranav2021} and \cite{Biagetti2022}. Gaussianity of cosmic microwave background radiation was studied with persistence diagrams in \cite{Cole2018}.

One of the most prominent applications of persistent homology in recent years is in our opinion the work by \cite{CisewskiKehe2022}, in which the authors directly develop the idea of discriminating power of persistence diagrams for cosmological models. They analyzed several metrics for computing the distance between persistence diagrams to check whether it is possible to tell apart the statistical difference between the persistence homologies of cold and warm dark matter universes. This, in principle, coincides with the idea of the present work, which is to utilize the Wasserstein distance for the analysis of the Cosmic Web and to monitor its sensitivity to variations of the $\sigma_8$ cosmological parameter, i.e. the present root-mean-square matter fluctuation averaged over a sphere of radius 8 Mpc comoving (see also \citealt{Biagetti2021} where the possibility of persistence homology methods to resolve between the impact of different $\sigma_8$ and non-Gaussianity parameter on Cosmic web was shown).

The graph-theoretical approach, known also as the complex network approach, discovers many aspects of the Cosmic Web (see, for example, an overview in \citealt{Tsizh2020}), but it still poses a significant disadvantage when applied to graph-like data.
It consists in the critical dependence of the constructed graph on linking length distance, a distance under which the two halos or galaxies are considered connected. The persistent homology is one of the possible non-parametric ways to explore the data, it liberates the explorer from disadvantage of casting pre-defined scale onto the analysis.

We present our work as follows. In  Sec.\ref{sec:homology} we present the core idea of persistence homology and convey the definition of Wasserstein distance two between persistence diagrams. In Sec.\ref{sec:simul} we describe the simulation we used for analysis and we formulate the hypothesis to test. In Sec.\ref{sec:results} we present our main results. In Sec. \ref{sec:discussion} we discuss limitations of our findings and the possibility of applying the new tool to the observational catalogs of galaxies. In the last section, Sec.\ref{sec:conclusions} we summarize the paper. 

\section{Persistent homology}
\label{sec:homology}
A number of computational problems in theoretical physics can be reduced to comparing points or clouds of points, be these points measured or calculated positions of real physical objects (e.g. stars, galaxies, gas molecules) or even states of a certain system in its configuration space. It turns out, there is no straightforward notion of the distance between two point clouds: considering each individual point leads to ``informational overflow,'' ambiguity, and in most cases will be too sensitive to small changes in the positions of points.  To solve the problem, one can avoid considering individual points, but extract instead certain generalized information about their distribution that will comprise a manageable amount of numbers and then look for some notion of distance for these sets of numbers. 

One of the possible approaches is to endow each point with a sphere of radius $r$ centered at the point and consider the union of all these balls as a manifold. This manifold will have certain topological properties that can be numerically represented and used further. As such, one uses Betti numbers\,--- ranks of appropriate homology groups of the given manifold. For simplicity, one can think of $0$-Betti number as a number of connected components, $1$- as a number of one-dimensional or ``circular'' holes, $2$- as a number of two-dimensional ``voids'' or ``cavities'' and so on. This notion is quite convenient for the problem under consideration as we can associate (unfortunately a bit indirectly) Betti numbers with cosmological structures: $0$-th with clusters, $1$-th with cycles and tunnels formed by cosmological filaments, and $2$ with cosmological voids formed by cosmic sheets.
Moreover, representation through Betti numbers is quite stable with respect to translation, rotation, and small variation of points' positions which is a desired property\footnote{In the code we extract topological features with so-called $\alpha$-complex without any restriction on the parameter $\alpha$. One may be concerned that it is defined in terms of simplices and Delaunay triangulation but not intersecting balls. Though, without any restriction on $\alpha$ it is equivalent to the \u{C}ech complex (but much smaller that is beneficial for calculation) and \u{C}ech complex in its turn is homotopy equivalent to the ``intersecting spheres'' point of view by the classic ``Nerve theorem'' \citep[][]{alexandroff1928}. Thus both views are equivalent, but ``intersecting spheres'' are easier to grasp.}. 

However, the choice of the  radius $r$ to use is debatable.
As a solution, one can consider all possible values of $r$ and track topological features that generate Betti numbers as the manifold changes with the change of $r$. Hence the name ``persistent homology'': each topological feature persists within a certain range of the parameter $r$ and Betti numbers change when features are ``born'' or ``die''. The result can be represented as a plot $r$ vs Betti numbers, or we can map ``birth'' and ``death'' of each individual topological feature in a form of a barcode diagram or persistence diagram. Interestingly, there is a notion of distance between persistence diagrams in a strict mathematical sense called ``Wasserstein distance'' and its partial case ``bottleneck distance.''
\begin{figure}
\includegraphics[width=0.95\columnwidth]{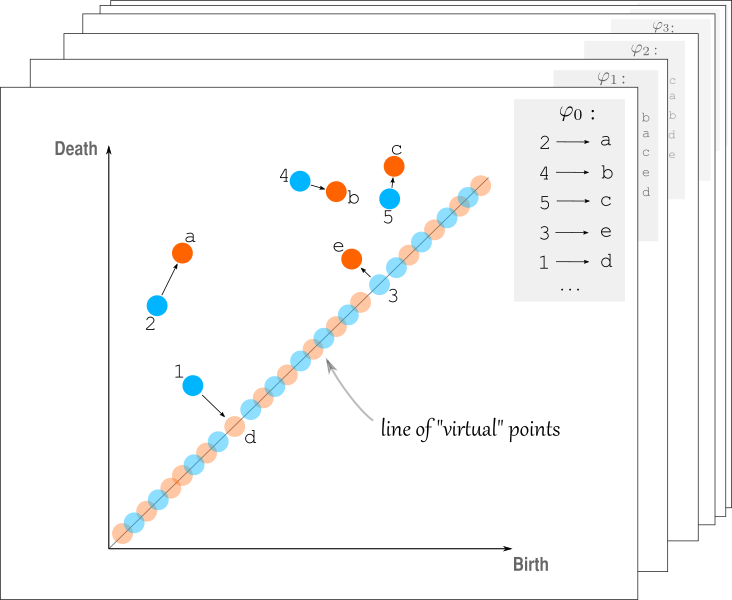}
\caption{
Maps $\varphi$ between persistence intervals as in  \eqref{eq_wasserstein}. 
Blue and red dots represent persistence intervals calculated from two different point clouds, their $XY$ coordinates represent the endpoints (``births'' and ``deaths'') of corresponding persistence intervals.  
}
\label{fig_leftover}
\end{figure}

\begin{figure*}
\begin{center}
\includegraphics[width=0.95\textwidth]{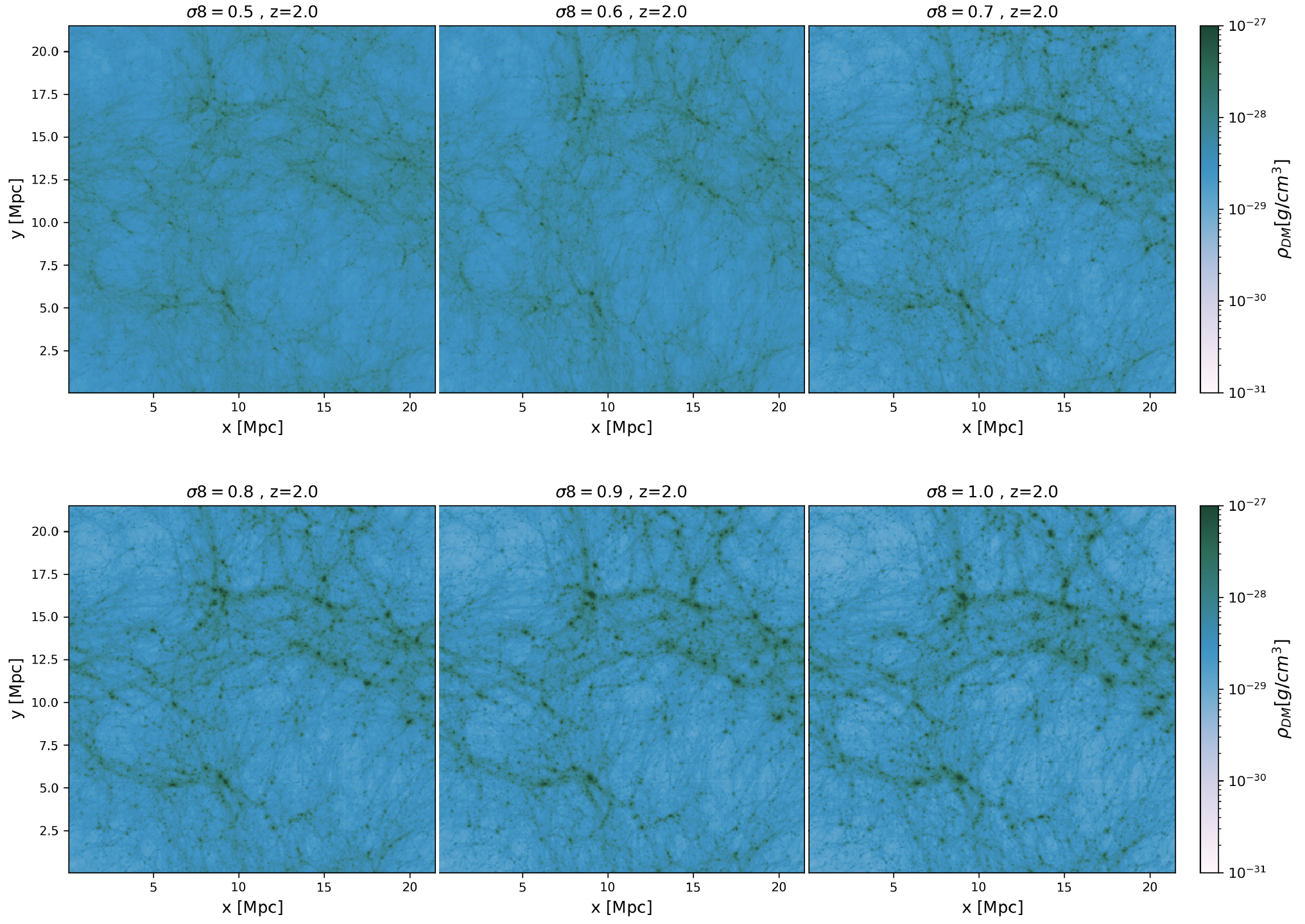}
\end{center}
\caption{
Projected maps of the average dark matter density along the full $42.5 \rm ~Mpc$ line of sight, for all variations of $\sigma_8$ for one of our ten randomly extracted initial conditions, for the epoch of $z=2.0$.
}
\label{map1}
\end{figure*}

\begin{figure*}
\begin{center}
\includegraphics[width=0.95\textwidth]{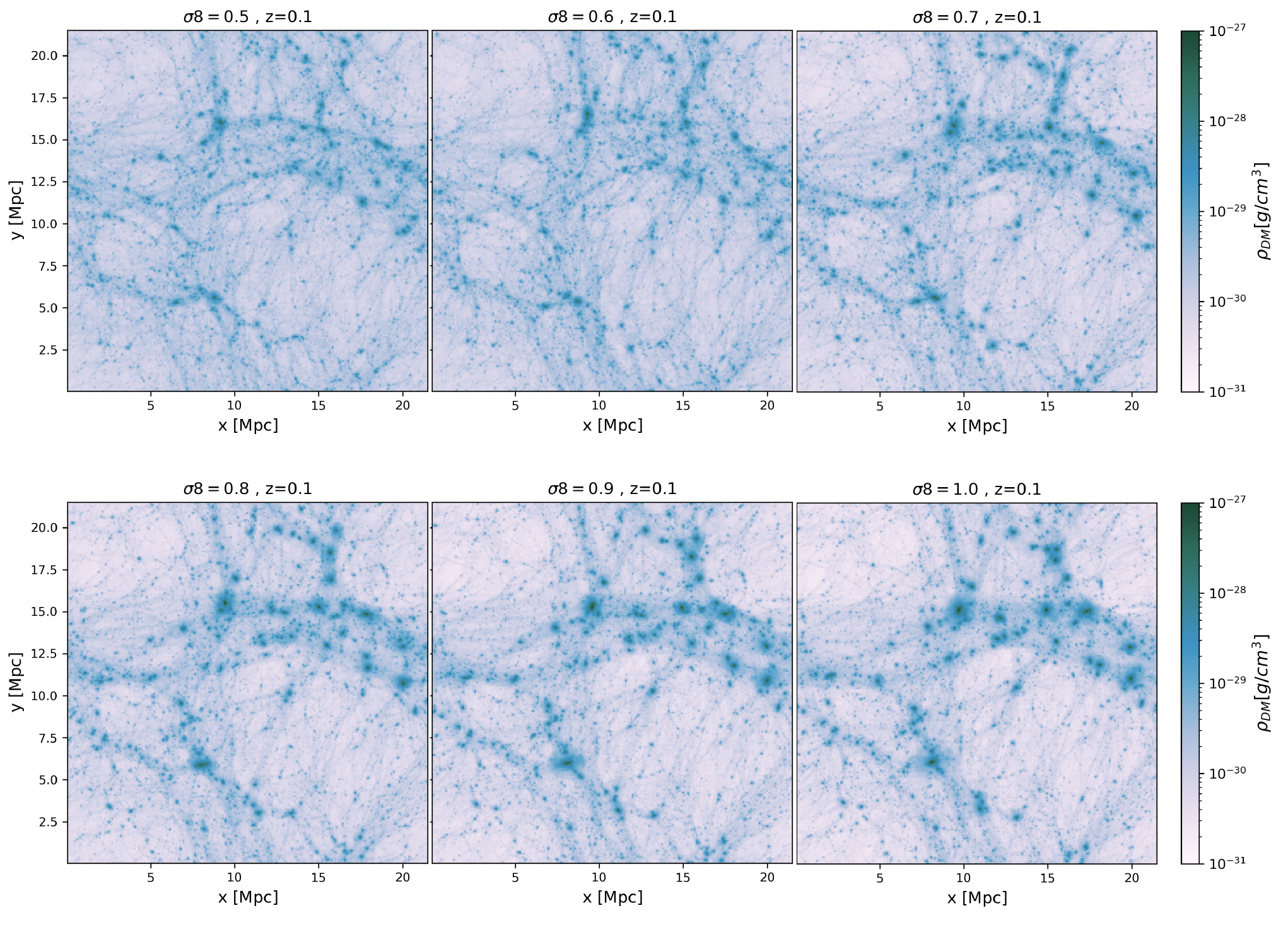}
\end{center}
\caption{
Projected maps of average dark matter density along the full $42.5 \rm ~Mpc$  line of sight, for the same model used in Fig.\ref{map1}, for the final epoch of  $z=0.1$.
}
\label{map2}
\end{figure*}

\subsection{Wasserstein and Bottleneck Distances}
\label{subsec:distance}
In probability theory and computational topology there is a widely used notion of $p$-Wasserstain distance \citep[][]{Edelsbrunner2010}
\begin{equation}
\label{eq_wasserstein}    
W_p(X; Y) = 
\left( 
    \inf_{\varphi: X \rightarrow Y} 
       \sum_{x \in X}
          \|x - \varphi(x)\|_{\infty}^p
\right)^{1/p}\hspace{-4.5mm},
\end{equation}
that provides distance between two multisets $X$, $Y$ (sets that allow multiple instances of any element). 
The idea of this distance is the following: first, we match elements of $X$ and $Y$ one-to-one (bijection $\varphi$) in a certain way (suppose it can be performed for now).
For each pair of elements $x \in X$ and $\varphi(x) = y \in Y$, we can calculate how much do they differ (the cost function) $\|x - \varphi(x)\|_{\infty}$ that is basically $L_\infty$ norm.
Adding up the $p$-th degrees $\|\cdot\|_\infty^p$ we get a notion of the difference between the whole multisets $X$ and $Y$ under the matching $\varphi:X \to Y$.
Taking the infimum over all possible bijections $\varphi$, we get the difference between multisets $X$ and $Y$ under the best matching possible effectively removing $\varphi$ from further consideration.
Now, taking the root of $p$-th order transforms the whole expressions into a well-defined distance that satisfies all axioms of metric \citep[][]{Figalli2021}.

Let us consider this distance with respect to our problem.
As we tracked how Betti numbers change with respect to radius $r$, the topological features were ``born'' and ``died''. Thus we got a persistence interval $[r_b;r_d]$ for each feature (can be represented as a point on a plane).
$X$ and $Y$ will now represent sets of such intervals calculated on different point clouds. 
The $L_\infty$ norm for two such intervals is defined as
$$
\left\|
   \left[r_b^A; r_d^A\right] - \left[r_b^B; r_d^B\right]
\right\|_\infty = 
\max\left(
   \left|r_b^A - r_b^B\right|, \left|r_d^A - r_d^B\right|
\right),
$$
where $r_b^A$, $r_d^A$, $r_b^B$, $r_d^B$ are "birth" and "death" radii. Plugging it into \eqref{eq_wasserstein} we get the desired expression.

One more adjustment in the procedure is necessary since the persistence diagrams consist of finitely many points above the diagonal that can vary in number depending on the initial point cloud they were generated from. To this finite multiset, we add the infinitely many points on the diagonal, each with infinite multiplicity \cite{Edelsbrunner2010} (``virtual points'') and allow matching with them. These extra points are not essential to the diagram, but their presence allows us to find bijection $\varphi$ even in cases there are different numbers of persistence intervals in $X$ and $Y$ as shown in figure~\ref{fig_leftover}.
One may note that taking infimum over all possible $\varphi$ we will always end up in a situation when most of the diagonal points are mapped to the diagonal points with the same coordinates thus adding $0$ (i.e. smallest possible $||x-\varphi(x)||_\infty^p$) to the total cost function. Only the ones matched with the off-diagonal points will add to the total cost.

The bottleneck distance is the Wasserstein distance, with parameter $p \to \infty$. 
Finding the appropriate limit, it can be shown that \citep[][]{Edelsbrunner2010}
\begin{equation}
\label{eq_bottleneck}
W_\infty(X; Y) =
\inf_{\varphi: X \rightarrow Y} \sup_{x \in X} \|x- \varphi(x)\|_\infty.
\end{equation}

The precursor of the modern notion of bottleneck distance was probably first introduced by Patrizio Frosini in 1990 \citep[][]{Frosini1990}.
\begin{figure*}
\begin{center}
\includegraphics[width=0.32\textwidth]{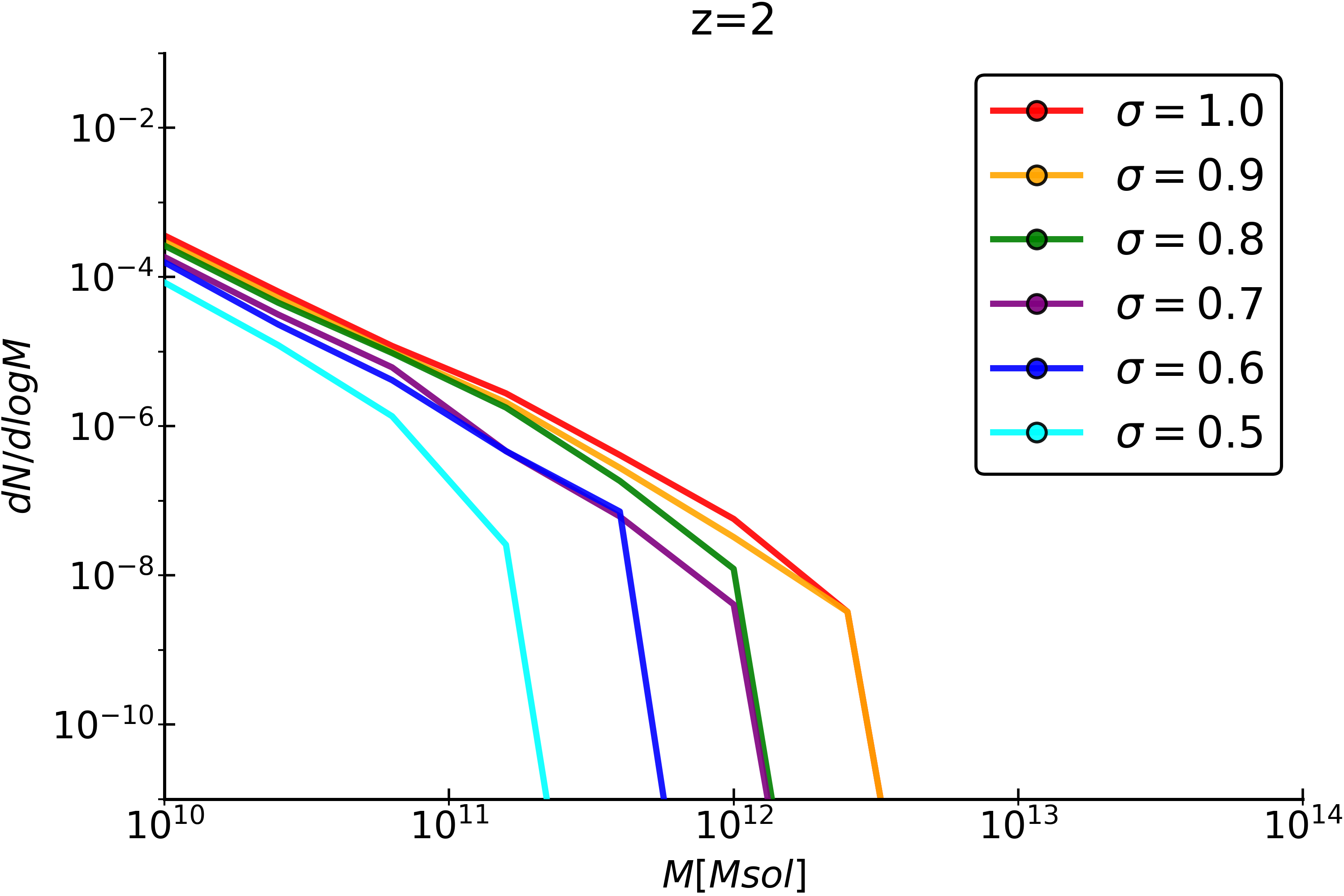}
\includegraphics[width=0.32\textwidth]{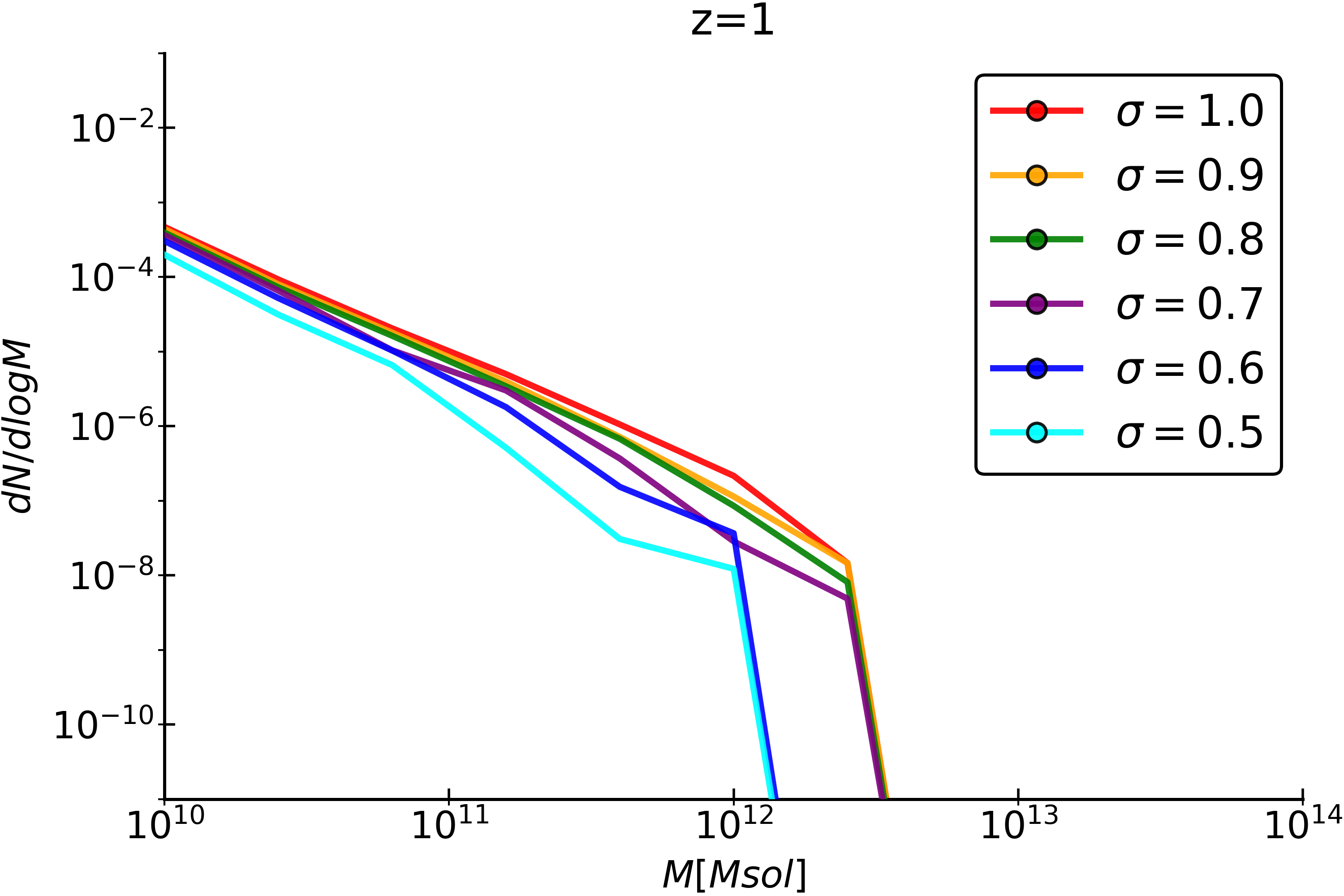}
\includegraphics[width=0.32\textwidth]{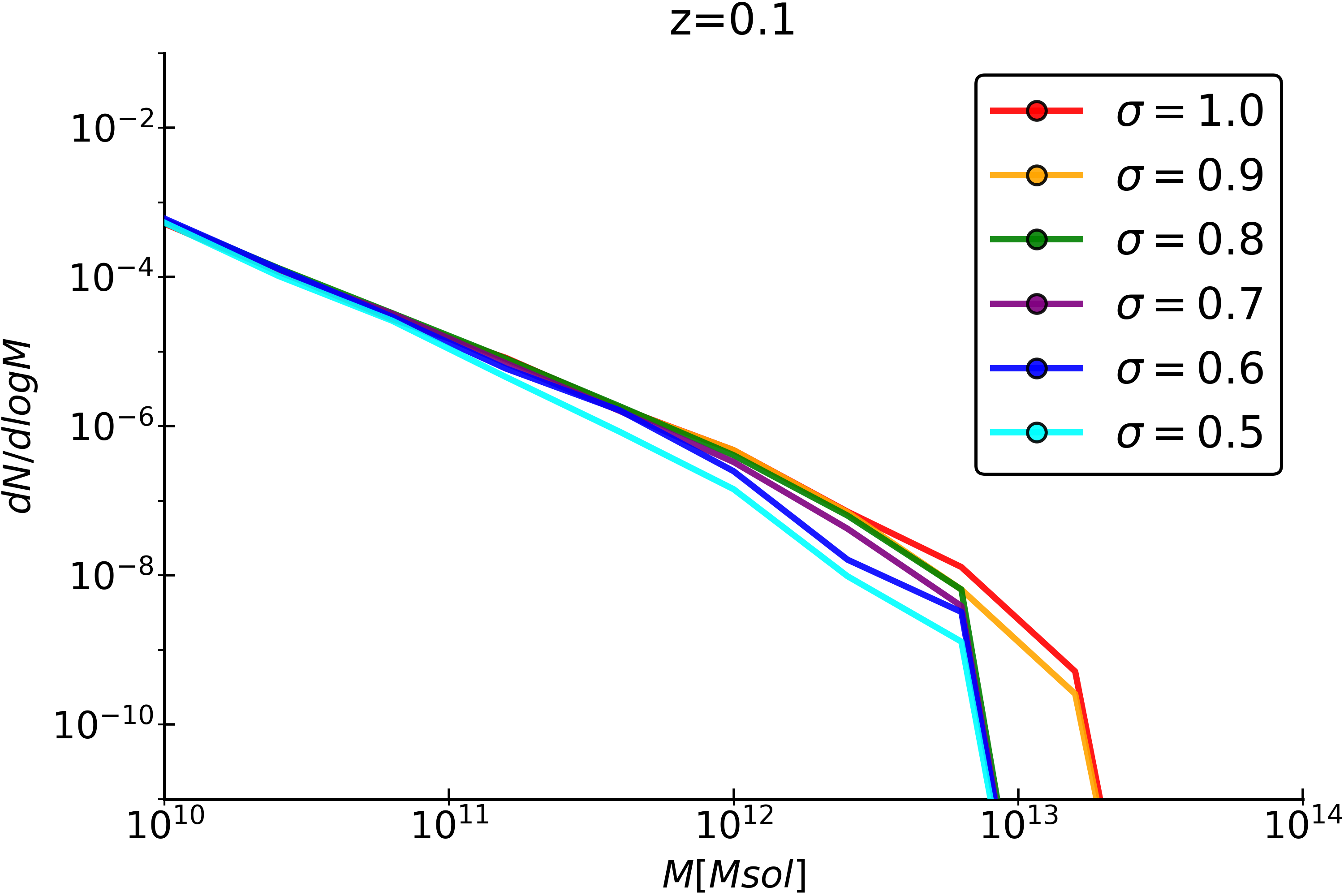}
\end{center}
\caption{
Halo mass functions (normalized to a comoving volume of $\rm 1 ~Mpc^3$) for all halos in our suite of simulations, for three different redshifts and for the six different simulated values of $\sigma_8$ (solid lines).
}
\label{halos}
\end{figure*}

\section{Cosmological simulations}
\label{sec:simul}
For this work, we analyzed a set of recent
cosmological simulations produced with the magneto-hydrodynamical code \texttt{enzo}\footnote{enzo-project.org}, applied to a suite of concordance $\Lambda$CDM  simulations of eight different cosmic volumes, for six different values of the $\sigma_8$ parameter. We used here a simple uniform resolution grid approach, sampling a comoving volume of $42.5^3 \rm Mpc^3$ with  $512^3$ cells (yielding a constant spatial resolution of $83.3~\rm kpc$/cell) and with $512^3$ dark matter particles (yielding a fixed mass resolution $m_{\rm dm}=6.19 \cdot 10^{7}M_{\odot}$).

These runs are part of a larger project, aiming at investigating the origin of cosmic magnetism through multiple resimulations of magnetic field seeding scenarios \citep[e.g.][and references therein]{va21magcow}. However, unlike what we did in most of the other projects along this line of research, here we fixed the initial magnetic field (assumed to have a primordial origin) and studied how the properties of simulated magnetic fields change with increasing initial amplitude of density perturbation ($\sigma_8$ parameter) and are also affected by cosmic variance.

Each simulation starts at $z=40$ from a "primordial" uniform volume-filling comoving magnetic field $B_0=0.1\ \mathrm{nG}$ for each magnetic field component. 

The simulation also includes the effect of radiative (equilibrium) cooling on baryon gas, assuming for simplicity a primordial chemical composition, and no additional sources of feedback. While these effects are not of primary importance for the study of homology presented here (which could have been done just using N-body simulations of dark matter, as usually done in the literature, the adoption of non-gravitational physics adds realism to the properties of the simulated gas network of the cosmic web, and it allows us to produce mock observables for these runs (subject of future works).

The cosmological parameters in this suite of simulations are kept  constant to the reference values of a flat $\Lambda$CDM cosmological model,  with $H_0 = 67.8$~km~s$^{-1}$~Mpc$^{-1}$, $\Omega_M = 0.308$, $\Omega_\Lambda =0.692$,  $\Omega_b = 0.0468$, {and $\sigma_8=0.815$} \citep{2016A&A...594A..13P}.  

With this setup, we generated a small suite of $48$ simulations, evolved from $z=40$ to $z=0.0$, for eight random variations of the initial phases of the matter and velocity distributions in the initial conditions (to produce eight independent random realizations of the same cosmology and gauge the effect of cosmic variance) as well as six different simulations with uniformly increasing $\sigma_8$, from $0.5$ to $1.0$.

We notice that compared to the recent work by \citet[][]{Bermejo2022} who first analyzed the simulated Cosmic Web using persistent homology, our suite of simulations investigates a much smaller (by a factor $\sim 800$) cosmic volume, while on the other hand, it provides a $\sim 10$ better mass resolution for halos, and it also allows us to monitor the effect of $\sigma_8$ and cosmic variance through the comparison of different resimulations.

Figures \ref{map1}-\ref{map2} give the visual example of the increased level of clustering of the dark matter component in one of our random seed extractions of the initial conditions, for the different $\sigma_8$ and at the early epoch of $z=2.0$, and towards the end of the simulation ($z=0.1$). 

The masses and positions of self-gravitating halos in all simulations are computed using the parallel Friend-of-friends (FOF) algorithm in \texttt{enzo}, imposing for all runs a linking length of $0.05$ cells and a minimum number of 50 dark matter particles for each halo, and computing their $M_{\rm 200}$ masses.

The dark matter mass distribution of halos in all our boxes  and for  all eight random variations of initial seeds is shown in Fig. \ref{halos}.
As for the projected masses of dark matter distribution, the effect of a decreasing $\sigma_8$ is seen, as expected, in a progressively decreased normalization of the measured mass function, which also leads to slower growth time of the most massive halos in the boxes.

On purpose, we do not present a detailed comparison with the theoretical expectations for the mass functions under the same cosmological models, because there are small, but not entirely negligible effects related to non-gravitational physics, which can affect the timing of halo formation as a function of $\sigma_8$, especially for the lowest values of it. 

Indeed, while all adopted non-gravitational effects are known to lead to negligible differences in the abundance of halos at low redshift (at least, for the range of baryonic physics and magnetic effects considered here), it is non-obvious to assess their impact at high redshift, and in the regime of initially low-density perturbations.
On one hand, a uniform magnetic field level can slow down the collapse of halos, by providing extra pressure to the gas \citep[e.g.][]{do99,2013ApJ...770...47K}, while on the other the impact or radiative gas cooling is that of accelerating the collapse of halos, leading to an increase up to a factor two with respect to non-radiative simulations at high redshift \citep[e.g.][]{2012MNRAS.423.2279C}.

Our suite of simulations has not been designed uniquely for this project, but also to compare with real radio observations of cosmic magnetism \citep[][]{2023MNRAS.518.2273C}, hence taking into account the presence of non-gravitational physics and its possible (secondary) effect on clustering.
In any case, it is interesting to study how the global effect of gravity and baryonic physics can alter the measured distribution of halos as a function of redshift and $\sigma_8$, which is the case that real observations have to face. 

More ad-hoc tests, in which single dependencies related to the input $\sigma_8$ can be tested and isolated through persistent homology will be considered in future work. 

\section{Results}
\label{sec:results}
\begin{figure*}
\begin{center}
\includegraphics[width=0.49\textwidth]{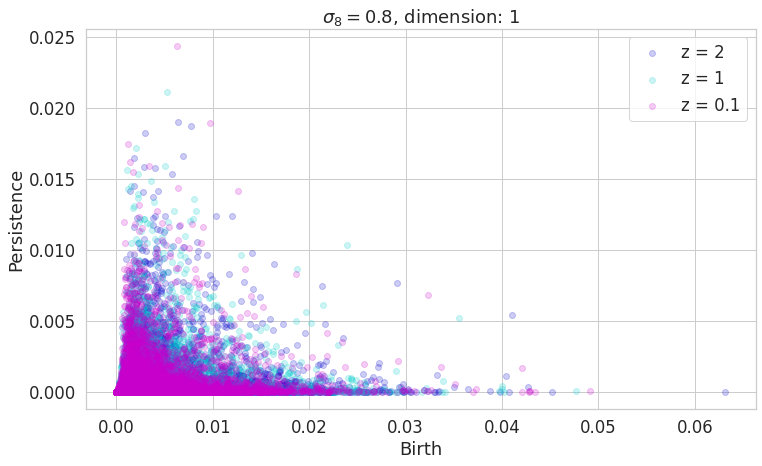}
\includegraphics[width=0.49\textwidth]{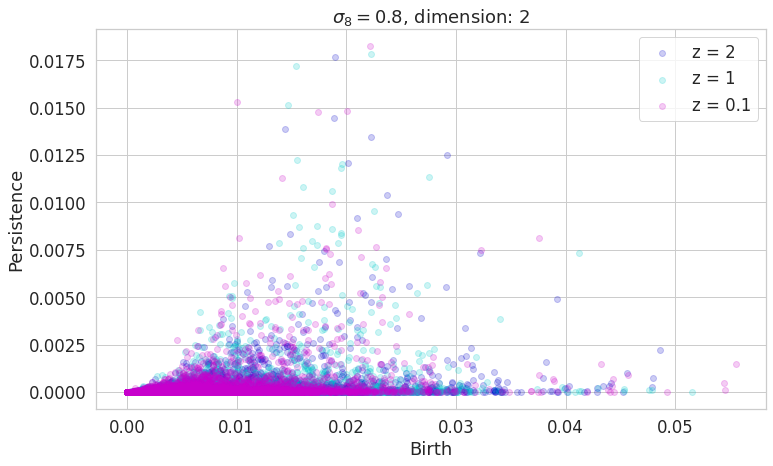}
\end{center}
\caption{Persistence-birth diagrams for 1-(left) and 2-(right) homologies for universes with $\sigma_8=8$ at different redshifts}
\label{persistencedeath}
\end{figure*}

First, we can make a qualitative comparison of our studies to others. Persistent homology was already applied to large-scale data. We can compare our (Fig. \ref{persistencedeath}) persistence (birth-persistence in this case) diagrams at different $z$ to those, for example in cite{Biagetti2021} (Fig. 6 in there). One can notice, that these images, in principle, have common tendencies. These are: 
\begin{itemize}
  \item persistence of the features on smaller scales at lower $z$. The points that correspond to $z=0.1$ are (on average) on the left to those which correspond to $z=1$, which are on the left of those, that correspond to $z=2$.
  \item 2-Homologies have wider and less steep distribution and on average they are born at a slightly larger radius.
\end{itemize}
These common tendencies show qualitative agreement between our study to those done before.

Now we will graphically present and elaborate on the numerical results. We have computed three types of defined above Wasserstein distances between each pair of simulated universes: the 1-Wasserstein, 2-Wasserstein, and $\infty$-Wasserstein (or the bottleneck) distance. The distances are measured between the "birth-death" diagrams of each pair. We compute distances for 0-dimensional, 1-dimensional, and 2-dimensional homology features (0-, 1-, and 2-homologies).
While 1- and 2-homologies can be treated as cycles of filaments and voids of the Cosmic Web respectively, 0-homologies are nothing else but connected components that start as separate points at $r=0$.

The distances, unless mentioned otherwise, are given in a range between 0 and 1, this is scaled values, where 1 corresponds to the length of the edge of a simulation cube, $42.5 \rm ~Mpc$.
We are interested in the dependence of these distances on the difference in $\sigma_8$ parameter between simulated universes, $\Delta\sigma_8$.
To take into account the impact of cosmic variance on our results, we have 8 simulations for each value of $\sigma_8$, yielding $8\cdot 6\cdot (8\cdot 6 - 1)/2 = 1128$ pairs of pointclouds, and, respectively distances calculated. We then average values of distances for each $\Delta\sigma_8$ value. The obvious disadvantage of such a way of averaging is, that we have a different number of points to the average for each value of $\Delta\sigma_8$, as the smaller $\Delta\sigma_8$ is, the larger number of pairs exist with such difference.
Indeed we have $6\cdot 8 \cdot (8-1) =  336$ points at $\Delta\sigma_8 = 0$ and only $8\cdot 8 = 64$ at $\Delta\sigma_8 = 0.5$. However, we have to reconcile with such a feature and rely on a large number of points even for $\Delta\sigma_8 = 0.5$.

\begin{figure*}
\centering
\begin{tabular}{|p{0.45\textwidth}|p{0.45\textwidth}|}
\hline
\includegraphics[width=0.45\textwidth]{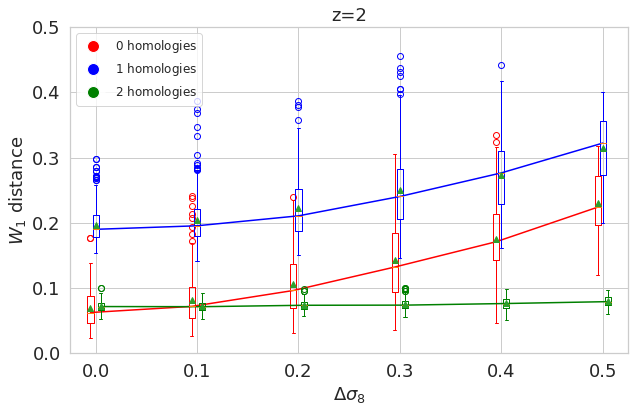} & \includegraphics[width=0.45\textwidth]{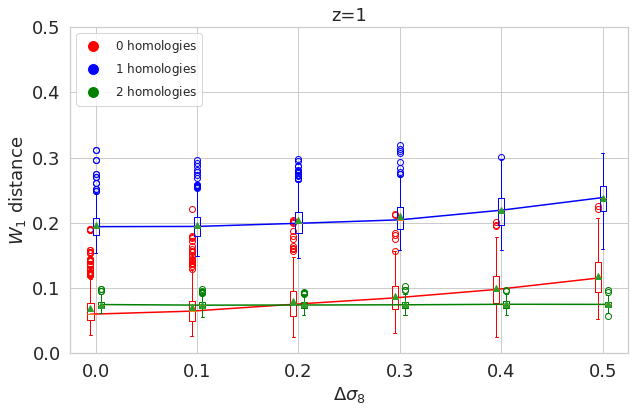} \\
\parbox[c]{\hsize}{\includegraphics[width=0.45\textwidth]{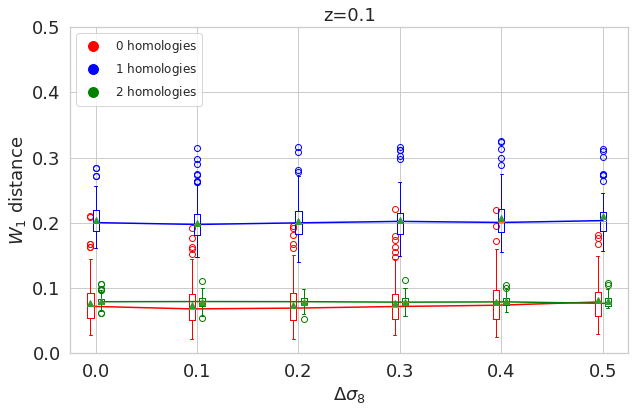}}  &
\caption{
Boxplots of $W_1$ distances at different $z$.
Each figure contains $3$ sets of data for homologies of different dimensions (color-coded).
Solid lines represent the general trend for median values. Additionally, mean values are shown as triangles.
Please note, data for $0$ and $2$ homologies was shifted a bit to the left and right to avoid boxplots overlapping.
}\label{fig_boxplots} \\
\hline
\end{tabular}
\end{figure*}

Results for different values of $z$ are represented in figure \ref{fig_boxplots}.
The $X$-axis shows the difference $\Delta\sigma_8$ between two simulations.
The computed Wasserstein distances between persistent diagrams for different simulations can be treated in this context as a realization of a random variable (e.g. different seeds contribute to the randomness of halo distribution).
Thus we represent results in form of ``box plots''\,--- a well-known type of diagram in statistics.
Its ``anatomy'' can be represented as follows:
\includegraphics[width=\columnwidth]{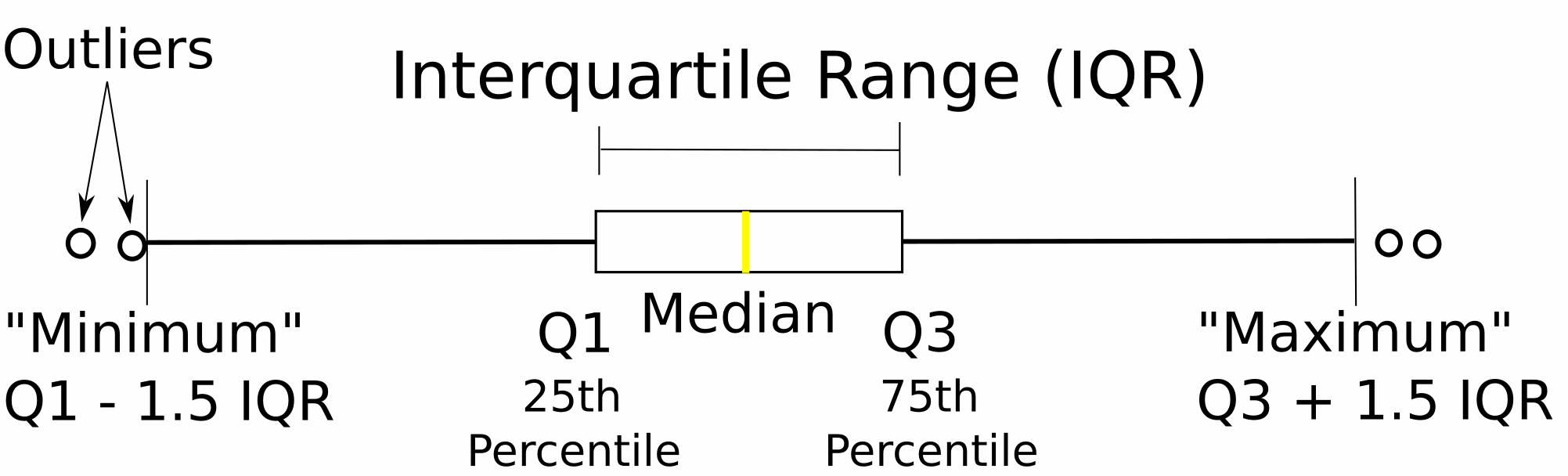}
Datapoints are sorted in ascending order and split into $4$ quartiles (Q1,...,Q4) each containing $25\%$ of the datapoints. The median shows where Q2 ends and Q3 starts. Calculations were performed on persistence diagrams of different order homologies separately (blue green and red colors for $0$-, $1$-, and $2$-homologies respectively), thus three curves in each plot.

Among all analyzed distances, 1-Wasserstein distances have the higher contrast between $\Delta\sigma_8 = 0$ and $\Delta\sigma_8 = 0.5$ for $z=2$ and $z=1$, this is why we select to show it on this figure. For $z=0.1$ $\infty$-Wasserstein distance has negligibly better contrast. Note, that each of the dimensions of homologies occupies its own niche of distances with the largest ones for 1-homologies. 

\begin{figure*}
\begin{center}
\includegraphics[width=0.49\textwidth]{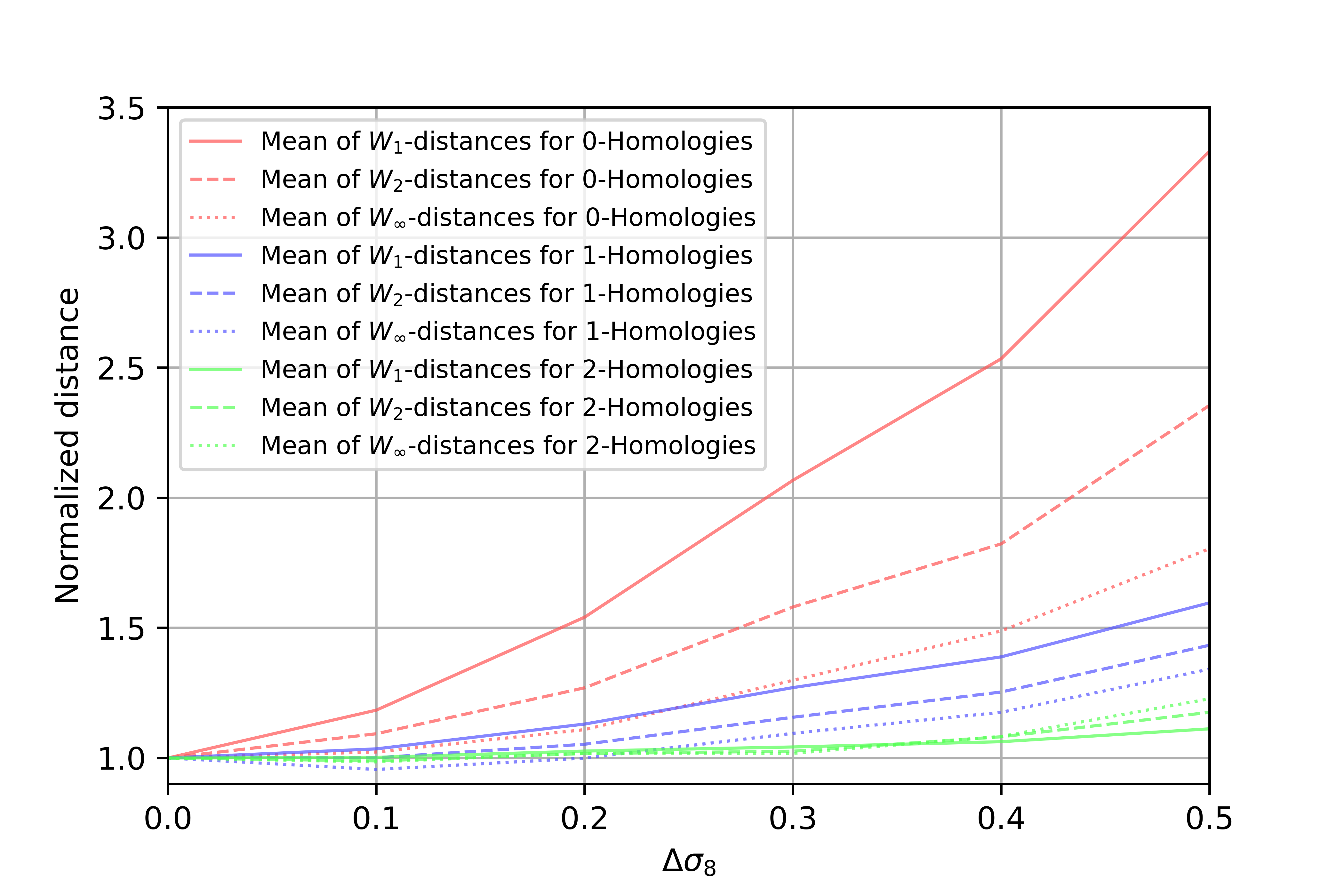}
\includegraphics[width=0.49\textwidth]{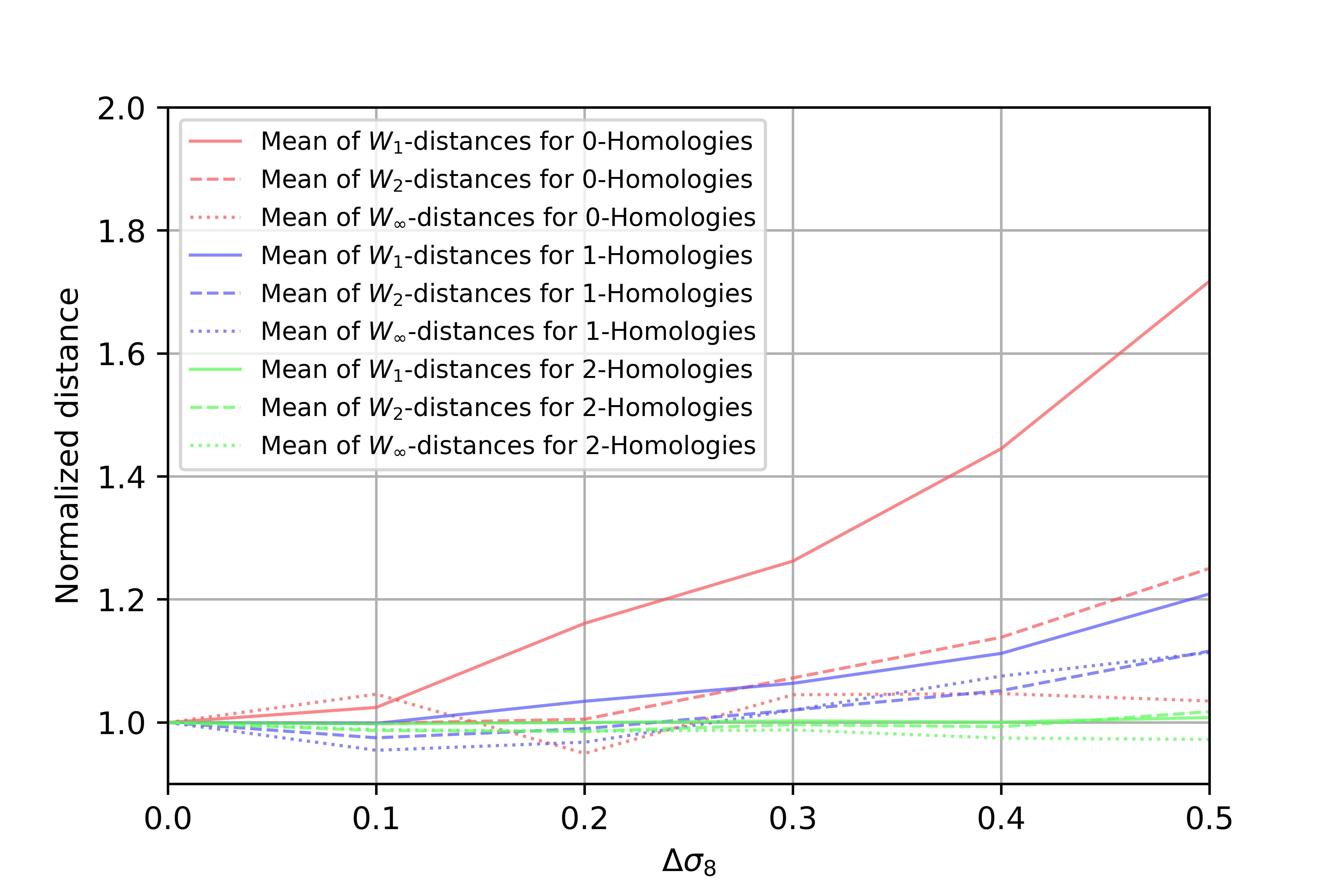}
\caption{Comparison of normalized distances changes with change of $\Delta \sigma_8$ at $z=2$ (left panel) and $z=1$ (right panel). Note, that y-axis scale in panels is different.}
\label{z21all}
\end{center}
\end{figure*}

\begin{figure*}
\begin{center}
\includegraphics[width=0.49\textwidth]
{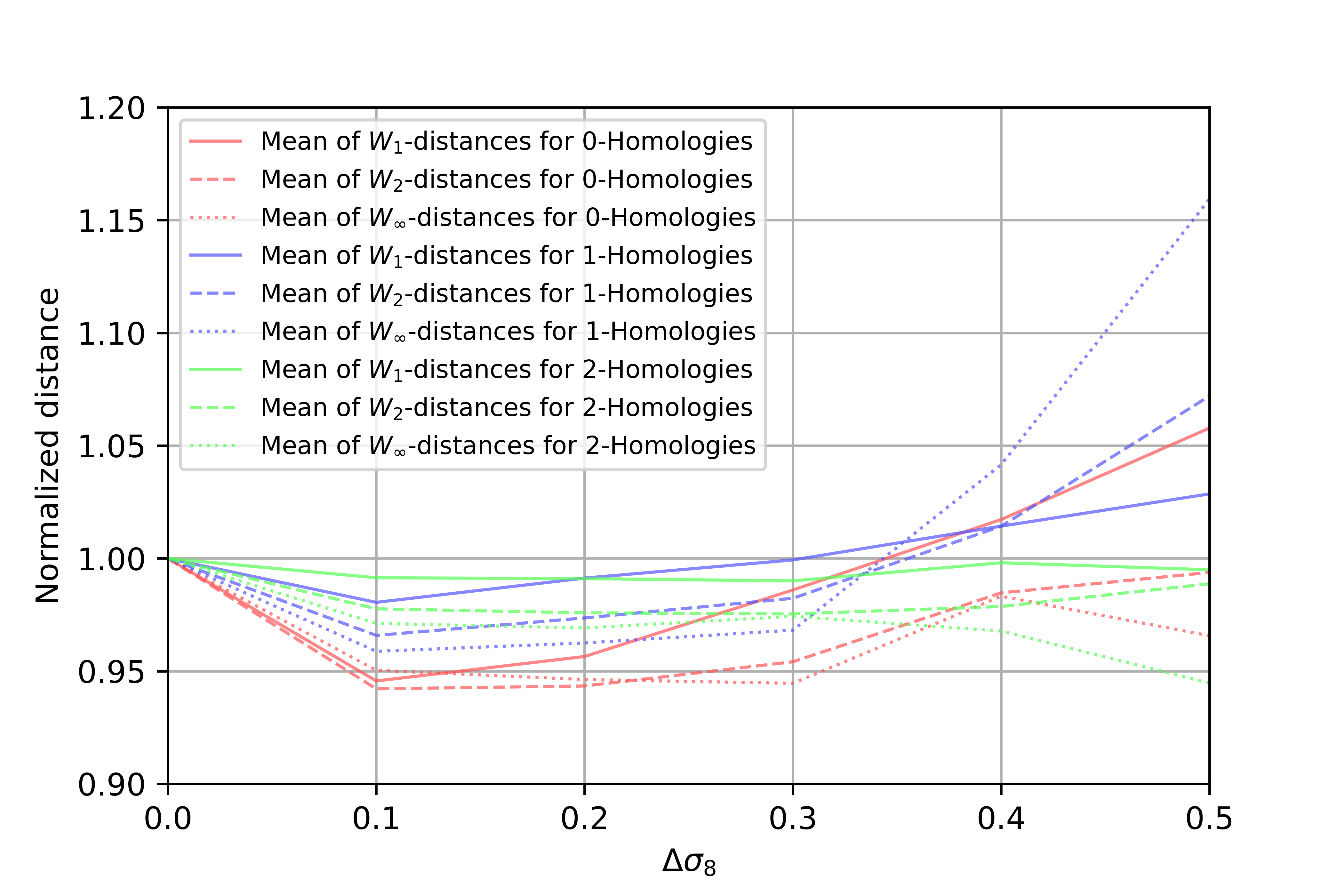}
\includegraphics[width=0.49\textwidth]
{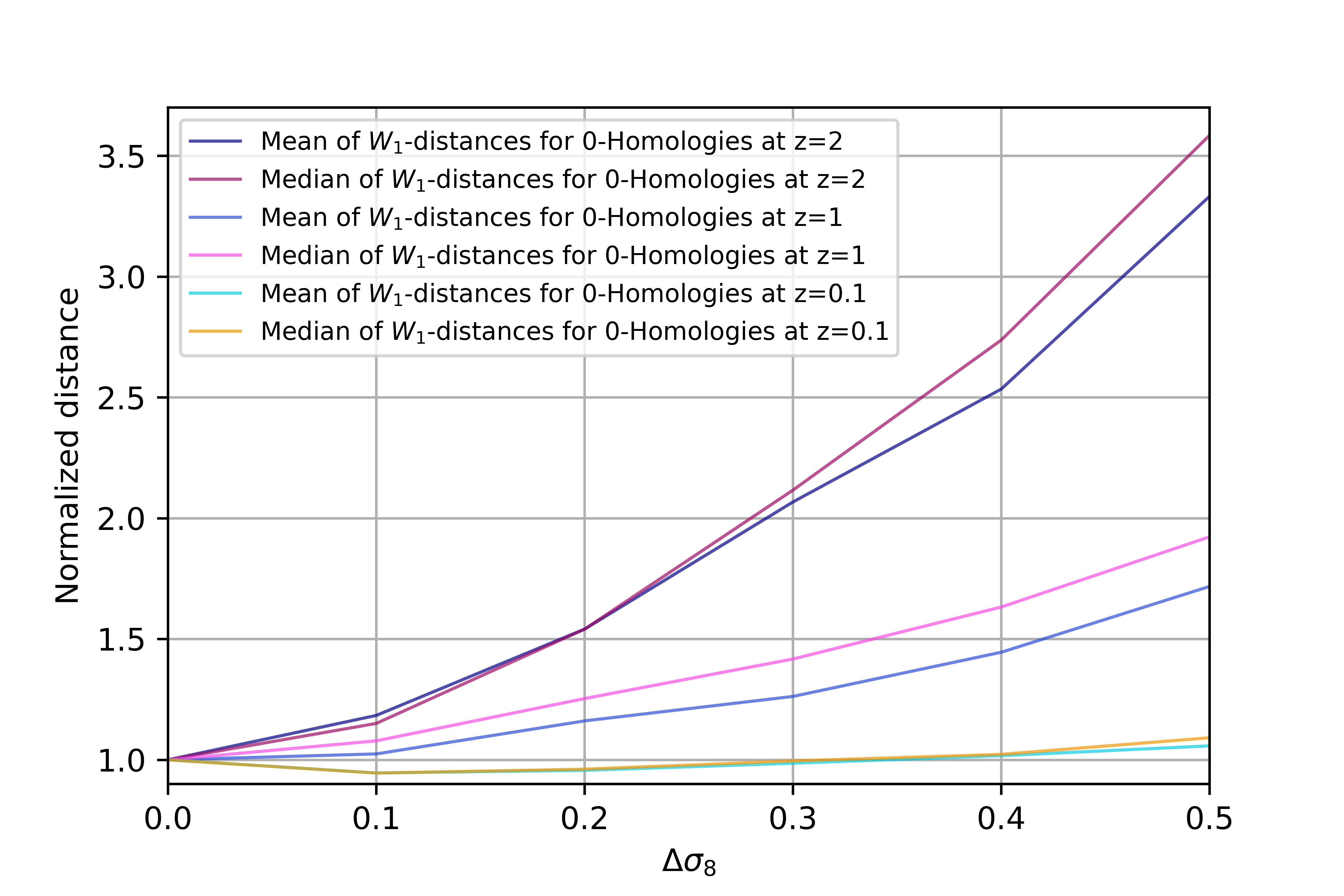}
\caption{Comparison of normalized distances changes with change of $\Delta \sigma_8$ at $z=0.1$ (left panel). Comparison of normalized $W_1$ distances at different $z$ (right panel)}
\label{zall}
\end{center}
\end{figure*}

In figures \ref{z21all} and \ref{zall} one can find the comparison of different variants of Wasserstein distances in distinguishing power between universes with varying $\sigma_8$ in normalized values. In these plots, each point of the line (which shows the averaged distance) is divided by its value at $\Delta\sigma_8 = 0$. In figure \ref{z21all} and the left panel of figure \ref{zall} the red lines represent the 0-homologies, the blue lines represent 1-homologies, and the green lines correspond to 2-homologies. Different styles of lines correspond to different types of distances (solid line to 1-Wasserstein, line-dotted to 2-Wasserstein, and dotted lines to $\infty$-Wasserstein distance). On the right panel of \ref{zall} one will find a comparison of distinguishing power of averages and medians of $W_1$ distance at different $z$. Medians might work slightly better in some cases. 

Let us deeper analyze these two sets of graphs. The first and obvious conclusion is that on average, Wasserstein distances reflect the size of the difference in $\sigma_8$ parameter between simulated universes. However, the distinguishing power of persistent homology decreases as redshift approaches 0. It appears that the networks of halos are better at breaking the degeneracy of $\Delta\sigma_8$ in the earlier Universe. The second conclusion is, that 0-dimensional homologies work better than 1-dimensional homologies, which, in turn, work better than two-dimensional ones. Persistent loops of filaments do a better job than persistent voids and sheets of matter, and comparing the persistence of separated components is the best of all. The averages and medians of distances for 0-homologies at $z=2$ increase their value by more than ~3.4, compared to ~1.7 for 1-homologies and 1.2 for 2-homologies. Meantime, at $z=0.1$ distances between universes with contrasting $\sigma_8$ become almost negligible. We can also note, that dissimilarities between different types of Wasserstein distances are minor, although they are still present. 

\begin{figure*}
\begin{center}
\includegraphics[width=0.49\textwidth]{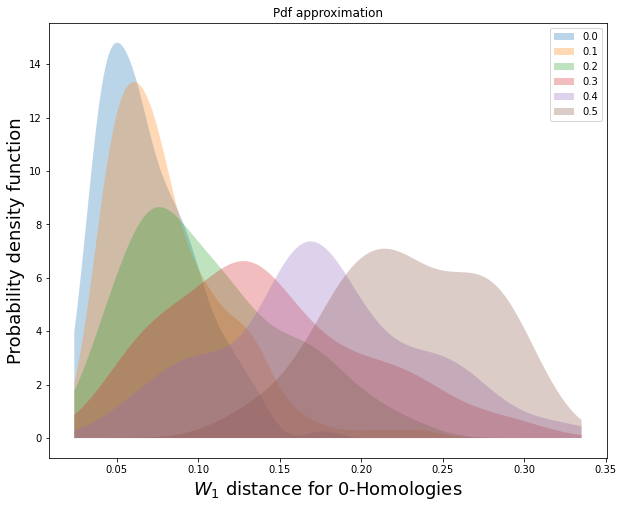}
\includegraphics[width=0.49\textwidth]{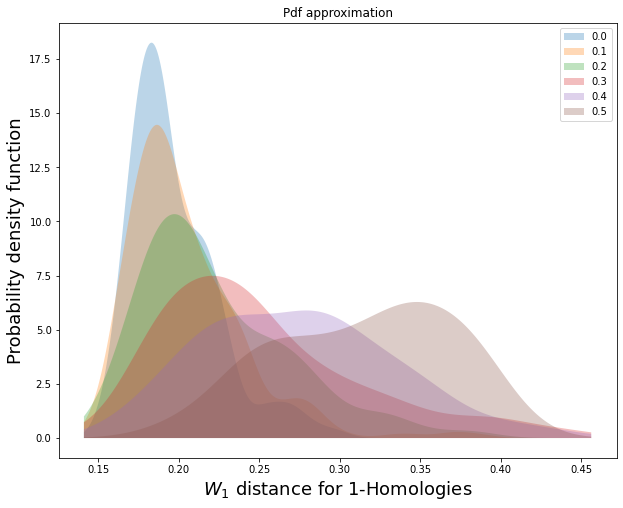}
\end{center}
\caption{Approximation of probability density functions for Wasserstein distances between simulated universes at z=2 for 0-Homologies (on the left) and 1-Homologies (on the right)}
\label{pdfs}
\end{figure*}

At the same time, the large width of the distribution of distances does not allow us to build a rigid statistical test, which would be able to set apart simulated universes with close values of $\sigma_8$. Indeed, as can be read from \ref{pdfs}, even for the best case of 1-Wasserstein distance of 0 and 1-homologies at $z=2$ the probability density functions of data points moderately overlap each other, and the situation worsens for low redshift data ($z=0.1$). In these graphs, we estimate the particle's distribution function with the kernel density estimation method (KDE). As a kernel, we used a Gaussian kernel with the width chosen by Silverman's rule.

\begin{figure*}
\begin{center}
\includegraphics[width=0.49\textwidth]{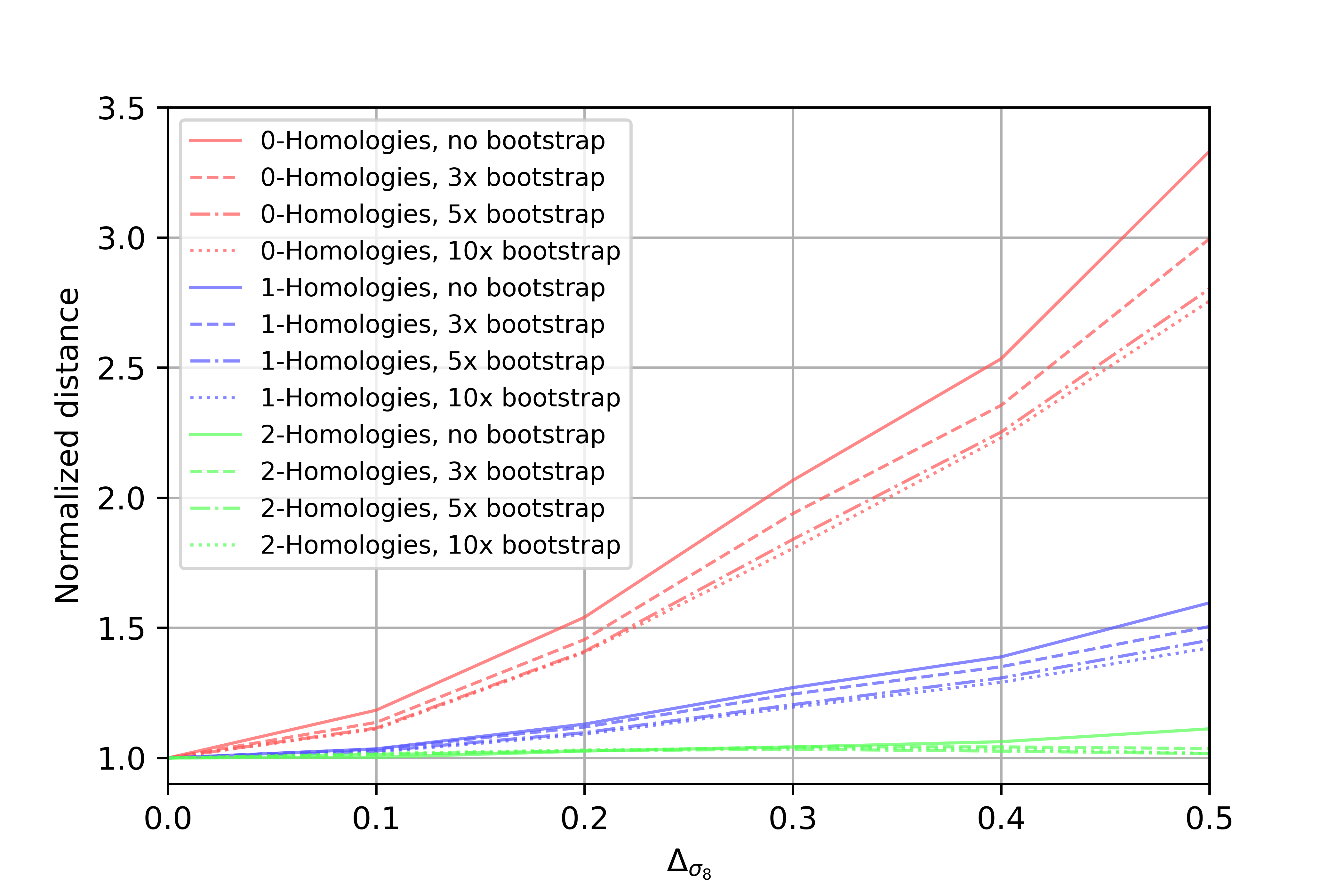}
\includegraphics[width=0.49\textwidth]{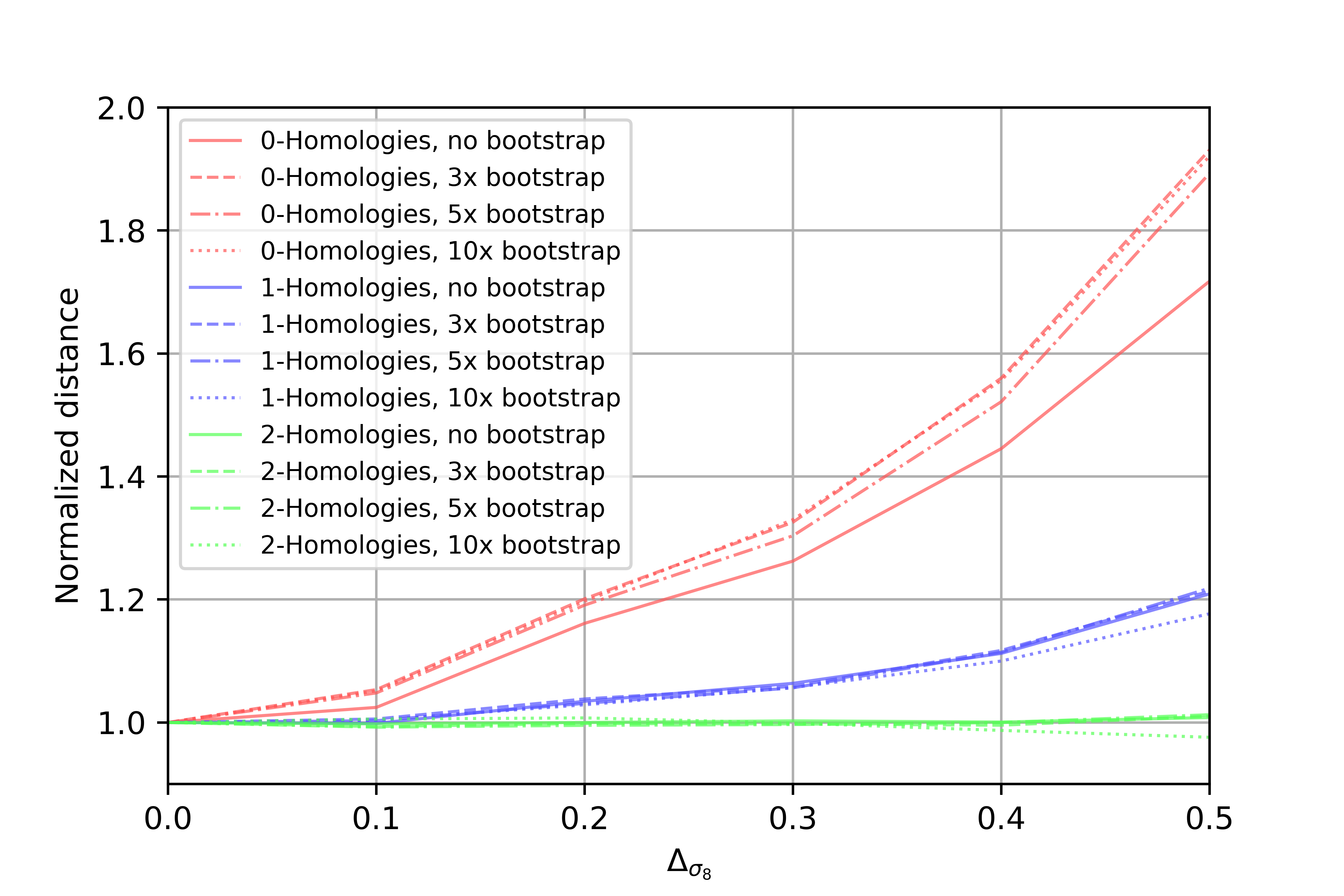}
\end{center}
\caption{$W_1$ distances at z=2 (left) and z=1 (right) with bootstrapped data}
\label{bootstrapped}
\end{figure*}

\begin{figure*}
\begin{center}
\includegraphics[width=0.49\textwidth]{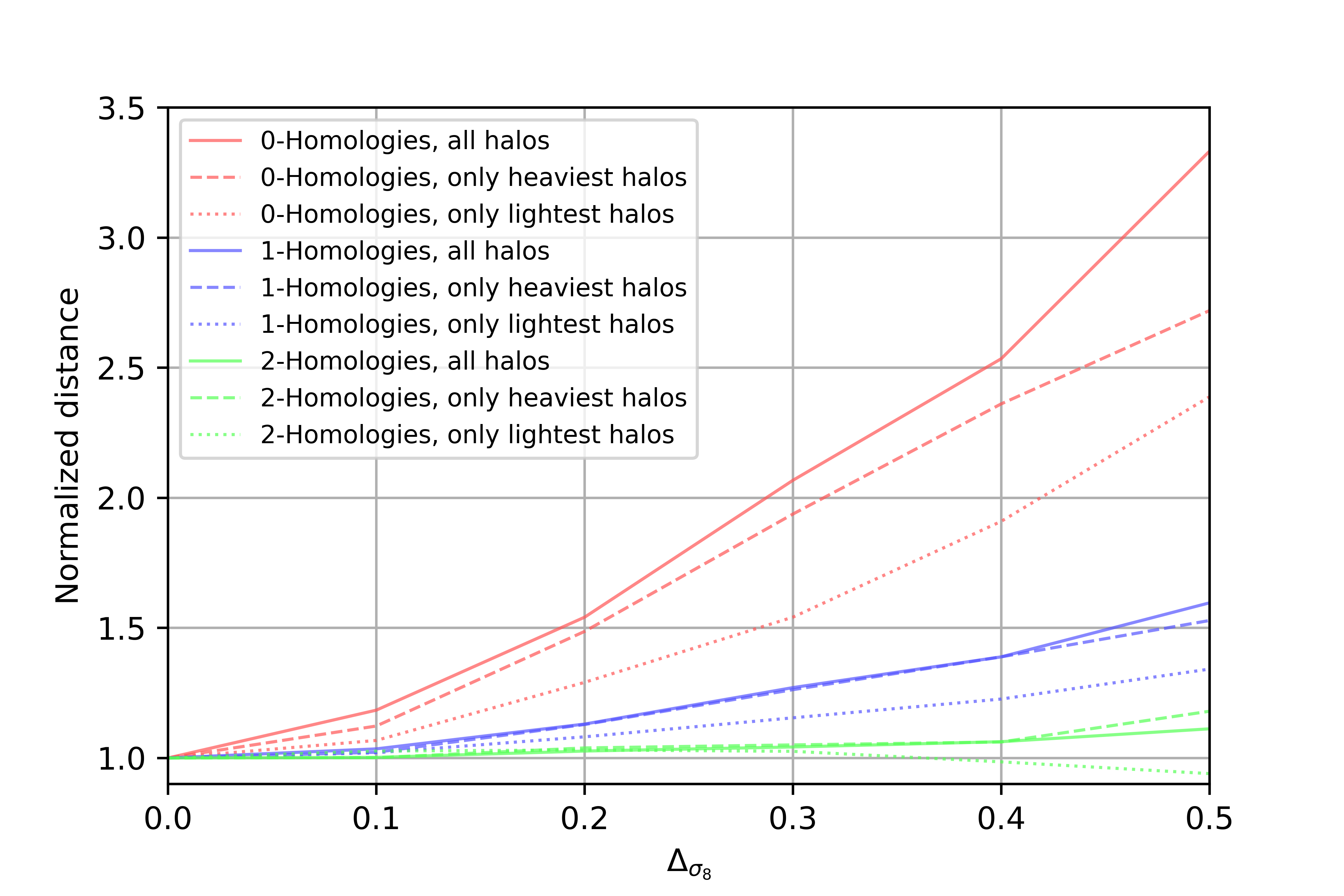}
\includegraphics[width=0.49\textwidth]{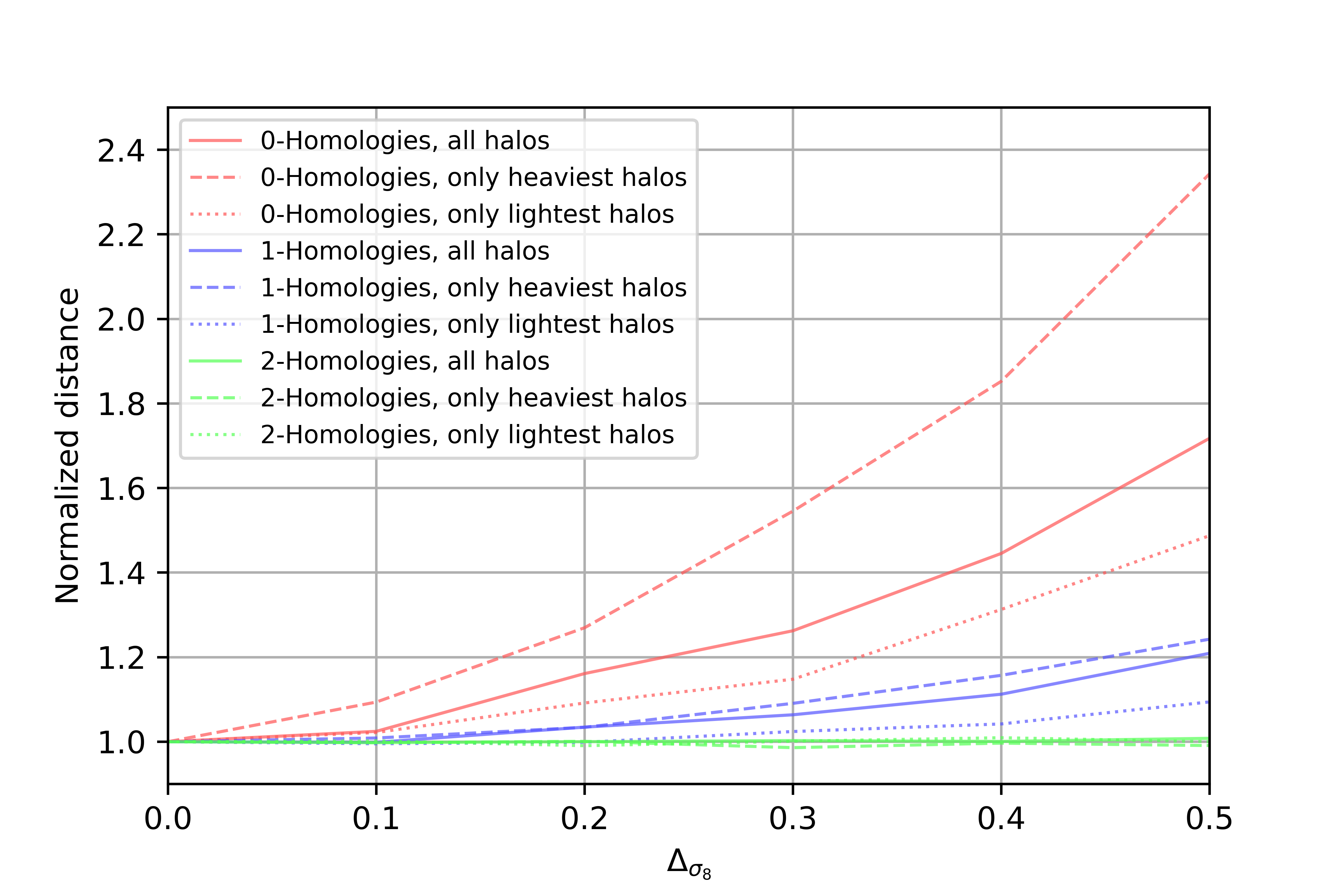}
\end{center}
\caption{$W_1$ distances at z=2 (left) and z=1 (right) with mass-filtered population of halos}
\label{massfiltered}
\end{figure*}

We have additionally checked, whether information about the differences is hidden in the "noise" of the large population of halos. We have calculated the same distances for "bootstrapped" data, where part of the halos was randomly deleted, averaging the overall result.
As can be seen for \ref{bootstrapped}, bootstrapping did not improve the results.
Likewise, the selective filtering out of the least or of the most massive halos did not produce a significant improvement of our results, and in Figure \ref{massfiltered} one can find and compare the averages of distances in populations, where only 10\% heaviest and 10\% of lightest halos are left.
The case of 0-homologies mass-filtering at $z=1$, where there is a considerable difference is rather an exception. This overall is a promising result: when applying this instrument to an even incomplete catalog of real galaxies, it is still possible to reconstruct the underlying topology of the network connecting them, in a robust enough way to constrain the value of $\sigma_8$. 

We have also tested whether the large span in a number of halos might cause large Wasserstein distances between universes instead of a difference in $\sigma_8$. To do this, we conduct all the same calculations on the data set, where the number of halos in all universes was equalized by truncation to a certain limit. We tested truncation (randomly throwing out excessive halos) at 800, 2000, and 4000 halos\footnote{Originally, for $z=2$ universes the span in the number of halos is $(849, 5264)$, for $z=1$ it is $(2373,7121)$, and for $z=0.1$ it is $(6353,10195)$}. If the number of halos in a universe was lower than this limit, we use bootstrapping (randomly repeatedly selecting the halos) to sample up this universe to the limit. In neither case, the result (sensitivity of $W_1$ to $\sigma_8$) wasn't considerably improved. In most cases, it was worsened and just in some, it was the same or slightly better. From this, we can abstemiously conclude that sampling only a small part of the population of the Cosmic web won't allow one to discern the nature of its homology features.


All of the computations were performed with \texttt{GUDHI} python library \citep[][]{Maria2014}, a specialized package for dealing with persistent homology. The plots were built with \texttt{Matplotlib} library of \texttt{Python}. All of our scripts and data can be found in open access\footnote{\url{https://github.com/mtsizh/bottleneck-distance-for-sigma8}}. It took a moderate amount of computational time to complete all of the calculations: a day on a modern desktop PC would be enough, which is an advantage of the considered method.

\section{Discussion}
\label{sec:discussion}
First of all, it shall be noted that, although persistent homology analysis requires no arbitrary choice of spatial scales, our results are still restricted by the limited spatial and mass scales sampled with our suite simulations (i.e. from $83 \rm ~kpc$ to $42.5 \rm ~Mpc$ comoving and from $6.1 \cdot 10^{7}$ to $\sim 10^{14} ~M_{\odot}$ for the dark matter component). However, theoretical works suggest that the self-organization of the Cosmic Web evolves only slowly (i.e. logarithmically) with scale \citep[e.g.][]{2007A&A...465...23S,2011CQGra..28p4003S}. The findings of \citet{Bermejo2022} confirm the logarithmic nature of this evolution also in terms of persistent homology and confirms the multiscale nature of the Cosmic web. This is why we expect our results to be possibly relevant to a wide range of spatial scales of the Cosmic Web. 

We would like to stress on the potential that persistent homology analysis has if applied to real data (either for future or even existing, surveys of galaxies). Unlike the more standard measurement of the mass function of galaxy halos as a cosmological probe, the persistent homology requires no knowledge of the halo masses (which can be increasingly more difficult to estimate for high redshift and/or small mass halos). Since only the three-dimensional position of galaxies, assumed to mark the location of dark matter halos, is needed to analyze the topology of the underlying matter distribution of the Cosmic Web in this approach, the only requirements are related to the accuracy of galaxy positions, in real surveys. Our additional tests also showed that this procedure is also robust against the removal of the least massive (or most-massive) halos in catalogs. 

As a comparison with real sky observations, we note that our analysis concerns eight independent realizations of the same cosmology (with variations of $\sigma_8$), which effectively cover a comoving volume of $85^3 \rm ~Mpc^3$. This corresponds to a projected field of view with size $\approx 2.7^{\circ}$ at $z=2$, $\approx 2.9^{\circ}$ at $z=1$, or $\approx 13.1^{\circ}$ at $z=0.1$, meaning that the area of sky which needs to be surveyed to test this technique is affordably small.
Moreover, the expected spectroscopic redshift uncertainty in future Euclid surveys is $\Delta z \approx 0.001(1+z)$ in the 0.7-1.8 redshift range \citep[e.g.][]{2021A&A...647A.117E}.
Although for a smaller field of view, similar redshift uncertainties also apply to existing surveys (e.g. $224 \rm ~arcmin^2$ for the VIMOS survey, \citealt{2018A&A...609A..84S}, and $1.7 \rm ~deg^2$ for z-COSMOS, \citealt{2007ApJS..172...70L}, to cite a few).
Significantly larger uncertainties, and restricted to a $z \leq 1$, are instead typically available in the SDSS survey \citep[e.g.][]{2019A&A...621A..26P}.
This corresponds to about 8kpc at z=2 and to about 5.5kpc at z=1, and it means that at both epochs the typical redshift uncertainty of future surveys is even smaller than the spatial resolution of the simulation we used to perform our homology analysis, and for this reason, we expect this technique to be, at least in principle, suitable to analyze the Cosmic Web even when realistic uncertainties in the redshift of galaxies are taken into account.
Moreover, it shall be stressed that even if in this work we have mainly focused on the effect of $\sigma_8$ on the growth of factor, which in our case also carries secondary effects related to the somewhat delayed growth of halos due to the non-linear interplay of other non-gravitational processes (such as primordial magnetic fields and reionization temperature background), in principle the same technique can be used to assess the impact of other physical parameters on the growth of structures, such as neutrino masses \citep[e.g.][]{castorina_2015}, warm dark matter and modified gravity models \citep[e.g.][]{2018MNRAS.473.3226B}, as well as other effects related to baryon physics \citep[e.g.][]{Shao2022}.

\section{Conclusions}
\label{sec:conclusions}
In this work, we presented new results from the application of persistent homology to study the multi-scale structure of the simulated Cosmic Web. Persistent homology is a powerful tool that allows us to decode the topological embedding of self-gravitating mass halos, and instead of focusing on the topology at a single (and arbitrary) length scale, it detects persistent topological features (measured through their Betti number) within the range of spatial scales in which they "born" and "die", as a result of increasing filtration. The Wasserstein distance resulting from this analysis (Sec.\ref{subsec:distance}) is thus a helpful and non-arbitrary measurement of the difference of typical scale characterizing the most persistent topological features of the Cosmic Web. Such difference, if induced by cosmological parameters (for example, $\sigma_8$), can be traced out and utilized to distinguish between different cosmologies.

In summary, our main conclusions are:
\begin{itemize}
  \item The Wasserstein distance shows the best discrimination power of $\sigma_8$ at the largest investigated redshift, $z=2$. The discrimination becomes increasingly worse moving towards low redshift (i.e. a little worse at $z=1$, and very low at $z=0.1$). 
  \item The persistence features of dimension 0 are much more sensitive to $\sigma_8$ than the features of dimension 1, which, in turn, are more sensitive than those of dimension 2. The distinguishing power decreases with the increasing dimensionality of features.
  \item The 1-Wasserstain distance shows the best results in comparing to 2-Wasserstein, or to the  Wasserstein distance ($\infty$-Wasserstein), in terms of discrimination of  $\sigma_8$, but differences are really small. 
  \item These findings are robust against bootstrapping, or the mass-filtering of the halos used for the network reconstruction. We also find that the median value of the above distance estimates, in general, is a little more informative than the mean of distances. 
  \item The distributions of Wasserstein distances for different values of $\Delta \sigma_8$ somewhat overlap one with another. Thus, the cosmic variance prevents us from the possibility to build a rigid statistical test of restricting the value of $\sigma_8$ by just comparing the persistence of homologies of Cosmic webs. 
  \item The selected spacial and mass ranges of our study allow us to carefully suggest possible similar exploration of persistent homology of the real Cosmic web, as modern galaxy catalogs possess the required size and resolution.
\end{itemize}
With all of this, we can predict, that the next steps in studies towards a deeper understanding of the topology nature of large-scale matter distribution would be exploring how other cosmological parameters shape the persistent homology of the Cosmic web, as well as comparing the homologies of the observable Cosmic web.

\section{Acknowledgements}
We would like to thank the Armed Forces of Ukraine for providing security to perform this work.
This work has become possible only because of the resilience and courage of the Ukrainian Army.
We would like to thank our anonymous reviewer for the helpful comments and suggestions, which have helped us to improve the quality of our work.
M.\,T. and F.\,V. acknowledge financial support from the Horizon 2020 program under the ERC Starting Grant \texttt{MAGCOW}, no. 714196. We also would like to thank the Italian MUR (Ministero Universit\'{a} e Ricerca), the Administration of University of Bologna, and especially L. Fortunato (Physics and Astronomy Department of Unibo) for the substantial administrative support, which also made this research possible.
M.\,T. was co-founded by the project of the Ministry of Education and Science of Ukraine “Modeling the luminosity of elements of the large-scale structure of the early Universe and the remnants of galactic supernovae and the observation of variable stars” (state registration number 0122U001834) and also would like to thank Dr. Veronika Romero and all the volunteers who help Ukrainian scientists during hard times.
In this work, we used the \texttt{enzo} code (\hyperlink{http://enzo-project.org}{http://enzo-project.org}), the product of a collaborative effort of scientists at many universities and national laboratories.
The authors gratefully acknowledge the Gauss Centre for Supercomputing e.V. (www.gauss-centre.eu) for supporting this project by providing computing time through the John von Neumann Institute for Computing (NIC) on the GCS Supercomputer JUWELS at J\"ulich Supercomputing Centre (JSC), under the project "radgalicm2".

\section{Data availability}
The data underlying this article are available in the \texttt{GitHub} repository at \url{https://github.com/mtsizh/bottleneck-distance-for-sigma8} and can be freely accessed by anyone. 
\bibliographystyle{mnras}
\bibliography{franco, maks, vitalii}

\label{lastpage}
\end{document}